\documentclass[aps,prl,twocolumn,amsmath,amssymb,floatfix]{revtex4-2}

\usepackage{graphicx}
\usepackage{dcolumn}
\usepackage{bm}
\usepackage{mciteplus}
\usepackage{braket}
\usepackage{upgreek}
\usepackage{amsmath}
\usepackage{amssymb}

\usepackage[svgnames]{xcolor}

\renewcommand{\section}[1]{\textit{#1}.---}
\usepackage[colorlinks=true,linkcolor=DarkBlue,urlcolor=DarkBlue,citecolor=FireBrick]{hyperref}

\begin{document}
\title{Instantons and the path to intermittency in turbulent flows}

\author{A.~Fuchs$^1$, C.~Herbert$^2$, J.~Rolland$^3$, M.~W\"achter$^1$, F.~Bouchet$^2$, J.~Peinke$^1$}
\affiliation{$^1$Institute of Physics and ForWind, University of Oldenburg, Küpkersweg 70, 26129 Oldenburg, Germany}
\affiliation{$^2$ Univ. Lyon, Ens de Lyon, Univ Claude Bernard, CNRS, Laboratoire de Physique, F-69364 Lyon, France}
\affiliation{$^3$ Univ. Lille, CNRS, ONERA, Arts et Métiers Institute of Technology, Centrale Lille, UMR 9014 - LMFL - Laboratoire de Mécanique des fluides de Lille - Kampé de Fériet, F-59000 Lille, France}
\date{\today}

\begin{abstract}
Processes leading to anomalous fluctuations in turbulent flows, referred to as intermittency, are still challenging. We consider cascade trajectories through scales as realizations of a stochastic Langevin process for which multiplicative noise is an intrinsic feature of the turbulent state. The trajectories are conditioned on their
entropy exchange. Such selected trajectories concentrate around an optimal path, called instanton, which is the minimum of an effective action. The action is derived from the Langevin equation, estimated from measured data. In particular instantons with negative entropy pinpoint the trajectories responsible for the emergence of non-Gaussian statistics at small-scales.
\end{abstract}

\maketitle
For turbulence, the phenomenon of intermittency~\cite{Frisch_1995} can be associated with the multifractal nature of the cascade~\cite{mandelbrot1977fractals}, 
but the physics leading to intermittency are still not fully understood.	
Here, we present a novel approach to this problem by a new characterization of the stochastic dynamics of velocity increments.

We adopt a description of the turbulent energy cascade by Markov processes~\cite{friedrich1997description}, i.e.  a \emph{Langevin equation} is used for the trajectories of velocity increment through scales.
We further interpret these trajectories as non-equilibrium thermodynamic processes~\cite{Seifert_2012,Nickelsen_2013,Reinke_2018,Peinke2018}. 
The \emph{entropy exchange}~\cite{Nickelsen_2013}, enables to distinguish between qualitatively different dynamics for increment cascade trajectories.
For negative values of the entropy exchange, there are preferential paths leading to the tails of the increment probability density function (PDF).

In many systems, a rare event follows a predictable path, known as \emph{instanton}, which is given as a minimizer of an appropriate action functional~\cite{ZinnJustinBook, FreidlinWentzellBook}.
Here, we use the instanton theory to compare experimentally measured and numerically generated trajectories.
The instanton approach has been used in several contexts (see Supplemental Material \cite{SM} \mbox{(Appendix A)})\nocite{Laurie_2015, Richardson22, kleinhans2005iterative, Kleinhans_2012, Nawroth2007, Risken, renner2002universality, Github}, like Burgers equation~\cite{Gurarie_1996,Balkovsky_1997}, random velocity field~\cite{Shraiman_1994, Falkovich_1996, Chertkov_1997, Balkovsky_1998}, 2D turbulence~\cite{Falkovich_2011}, shell models~\cite{Biferale_1999, Daumont_2000}, Lagrangian model of velocity gradient dynamics~\cite{Moriconi_2014, Grigorio_2017, Apolinario2019} or even geostrophic turbulence~\cite{Bouchet_2019,Simonnet_2021}, and rogue waves~\cite{Dematteis_2018, Dematteis_2019}.
A novelty of the present work is that we do not study the effect of external driving noise, but use an intrinsic noise of the cascade captured by an effective diffusion model of the cascade derived empirically from experimental data.
The turbulent state is thus based on intrinsic features without relying on the Navier-Stokes equation.
We then compute instantons defined as trajectories conditioned on entropy exchange.
The comparison of rare trajectories from experimental data and the learned effective stochastic dynamics is a very stringent test for the validity of the effective model. 
Based on this approach we show that a simplified version of our effective cascade model, corresponding to log-normal statistics~\cite{kolmogorov1962refinement, Oboukhov_1962}, does not capture the dynamics through scales correctly, while a small modification of this model does.


\section{Experimental and numerical data}
\label{sec:data}
From hotwire velocity measurements, $v(t)$, (component in direction of the mean flow) of fractal grid flow~\cite{Fuchs_2020}, we obtain the longitudinal velocity increments $u_r=v(t+\tau)-v(t)$, relying on the Taylor hypothesis~\cite{taylor1938spectrum} $r=-\tau\langle v\rangle$. 
The velocity increments $u_r$ are computed in units of
$\sigma_\infty=\sqrt{2}\sigma_v=3.5$\,m/s ($\sigma_v$ is the standard deviation of $v(t)$).

One classically studies \emph{structure functions}~\cite{Anselmet_1984} \mbox{$S_n(r)=\mathbb{E}[u_r^n]$} or PDFs~\cite{Castaing_1990} $P(u, r)=\mathbb{E}[\delta(u_r - u)]$
as a function of scale $r$.
Here, we consider the stochastic dynamics of velocity increments $u_r$ as they go through the cascade process.
As a natural description of the cascade we introduce the change of variable $s=-\ln (r/L)$,
from $s_i=0$ at the integral scale $L$ to a positive value $s_f=-\ln (\lambda/L)$ at the Taylor scale $\lambda$~\cite{friedrich1997description}.
\emph{Cascade trajectories} defined as $\left[u(\cdot)\right]=\{u_0,\dots,u_f\}$, with $u(s_i)=u_0$, $u(s_f)=u_f$
	are illustrated in Fig.~1 in the Supplemental Material \cite{SM} (Appendix B).
$\left[u(\cdot)\right]$ denotes the entire path instead of a distinct value $u_s$.

Previous studies have shown that the stochastic process $u_s$ can be considered as Markovian
~\cite{friedrich1997description, Marcq_1998, renner2001}, at least down to a scale close to the Taylor scale \mbox{($\Delta_{EM} \approx 0.9 \lambda$~\cite{Lueck2006markov, renner2001}).}
We model it as a diffusion process~\cite{GardinerBook} across scales
\begin{eqnarray}
	d u_s = D^{(1)}(u_s, s)ds + \sqrt{2D^{(2)}(u_s, s)} \circ dW_s,
	\label{eq:sde}
\end{eqnarray}
where $W$ is the Wiener process.
We use the Stratonovitch convention because we will consider in the sequel scale-reversed ($s\to s_f-s$) trajectories.

Two data sets of trajectories are generated by numerical integration of the stochastic differential equation~(\ref{eq:sde}) using a simple continuous diffusion model of the cascade
\begin{eqnarray}
	D^{(1)}(u, s) = -(\alpha+\gamma) u, \quad D^{(2)}(u, s) = \beta + \gamma u^2.
	\label{eq:betamodel}
\end{eqnarray}
The initial condition is a centered Gaussian distribution with variance $\sigma^2=0.082$, which fits well the experimental PDF at the integral scale.
The \emph{Kolmogorov-Obukhov} theory~\cite{kolmogorov1962refinement, Oboukhov_1962},  (\mbox{K62}), can be recovered with \mbox{$\alpha=(3+\mu)/9$,} \mbox{$\beta=0$}
and \mbox{$\gamma=\mu/18$}, where \mbox{$\mu=0.234$} is the intermittency parameter~\cite{friedrich1997description}.
In this case the SDE~(\ref{eq:sde}) is known as \emph{geometric Brownian motion} which has the analytical solution
\mbox{$u_s = u_0 \exp\left[-\left( \alpha+\gamma \right) s+\sqrt{2\gamma} \left( W_s-W_0 \right)\right]$}.

Based on eq.~(\ref{eq:betamodel}) the evolution of the PDF's and structure functions 
can be compared with experimental findings.
As shown in Supplemental Material \cite{SM} \mbox{(Appendix D)} for $\beta=0$ (K62) and $\beta=0.044$ (chosen empirically) the stochastic models correctly reproduce the statistics of increments including the fingerprint of intermittency expressed by the kurtosis, which can be further improved by scale depending $\alpha(s)$, $\beta(s)$ and $\gamma(s)$ (see Supplemental Material \cite{SM} (Appendix C).

\section{Action and entropy}
\begin{figure}
	\includegraphics[width=0.24\textwidth]{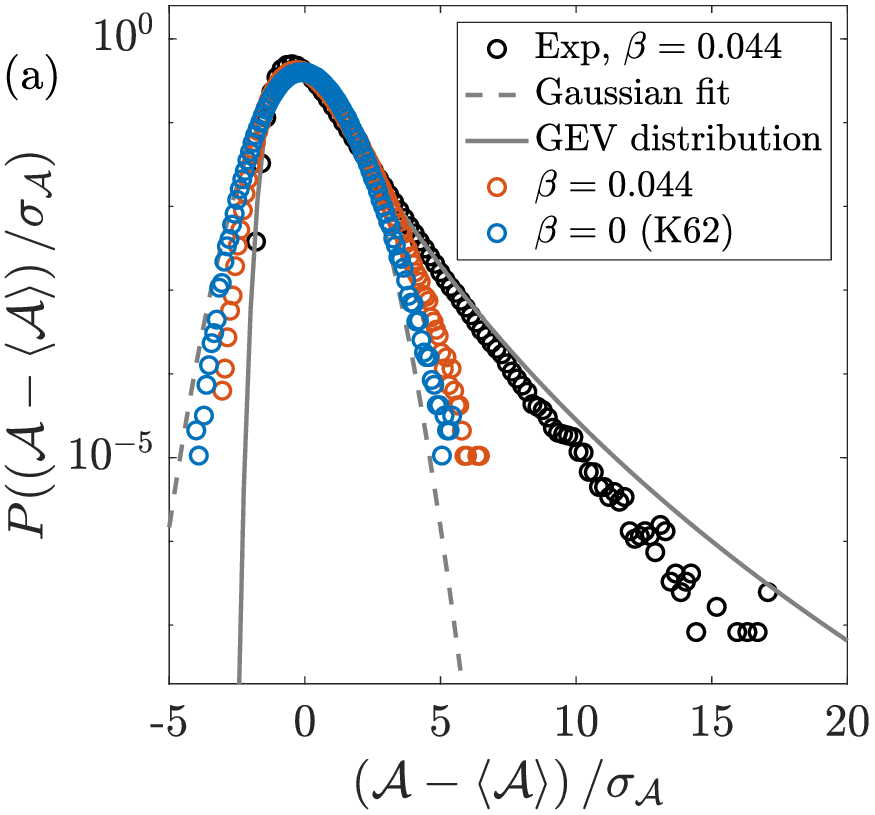}
	\includegraphics[width=0.235\textwidth]{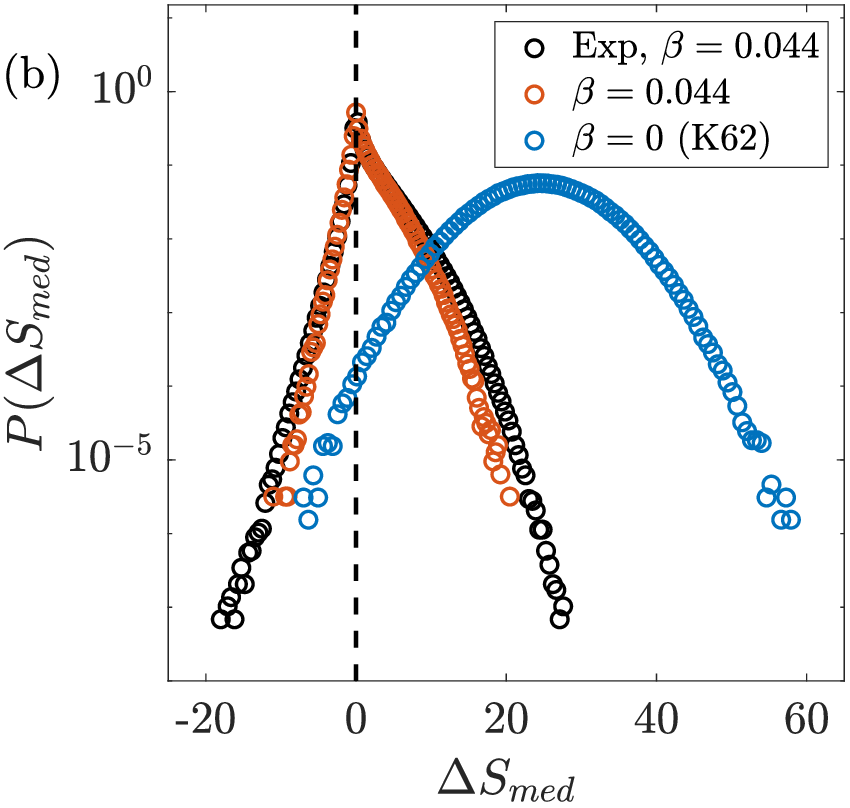}
	\caption{(a) PDF of normalized action estimated using experimentally acquired and numerically generated cascade trajectories. The grey dashed resp.\ solid line corresponds to a Gaussian fit and a generalized extreme value distribution fit to the experimental data, with the shape parameter $k = 0.1$. 
			(b) PDF of the entropy exchanged with the surrounding medium.}
	\label{fig:Action}
\end{figure}

Next we discuss the 
\emph{path integral formalism} cf.~\cite{ZinnJustinBook}.
The probability of a trajectory is proportional to
$exp\left[-\mathcal{A}\left[u(\cdot)\right]\right]$,
where $\mathcal{A}$ is the Onsager-Machlup action~\cite{onsager1953fluctuations,machlup1953fluctuations}.
The expectation value of any observable $\mathcal{O}$ can be computed as a weighted-integral over all possible trajectories \mbox{$\mathbb{E}[\mathcal{O}] = \frac{1}{\mathcal{Z}} \int \mathcal{D}\left[u(\cdot)\right] \mathcal{O}\left[u(\cdot)\right] e^{-\mathcal{A}\left[u(\cdot)\right]}$},
where $\mathcal{D}\left[u(\cdot)\right]$ is the measure over the space of trajectories and $\mathcal{Z}$ a normalization factor.
$\mathcal{A}$ depends on the discretization.
For 
~eq.~(\ref{eq:sde}), the action is
\begin{eqnarray}
  \mathcal{A}\left[u(\cdot)\right] = \int_{s_i}^{s_f} 	\left\lbrack\frac{\left(\dot{u}_s-D^{(1)} +D'^{(2)}/2 \right)^2}{4 D^{(2)}} + \frac{D'^{(1)}}{2} \right\rbrack ds,
	\label{eq:action}
\end{eqnarray}
interpreted in the Stratonovitch convention~\cite{Graham_1973, Arnold_2000, Lau_2007}.
Here we dropped the arguments of $D^{(1,2)}(u, s)$.
Figure~\ref{fig:Action}(a) shows the PDF of $\mathcal{A}$ estimated using eq.~(\ref{eq:action}), which is essentially Gaussian for the trajectories generated numerically ($\beta=0$, $\beta=0.044$) while the experimental data ($\beta=0.044$) is characterized by a heavy tail on the right.

Non-equilibrium stochastic thermodynamics~\cite{Seifert_2012} allows to associate with every trajectory a total entropy variation~\cite{Seifert_2005, Seifert_2012, Sekimoto_2010, Nickelsen_2013, Reinke_2018}, given by the sum \mbox{$\Delta S_{tot}\left[u(\cdot)\right] =\Delta S_{med}\left[u(\cdot)\right] +\Delta S_{sys}\left[u(\cdot)\right]$},
 which are related to heat, work and internal energy~\cite{Seifert_2005,Seifert_2012,Sekimoto_2010}.
The system entropy $\Delta S_{sys}\left[u(\cdot)\right]=-\ln{\left( \frac{P(u_f, s_f)}{P(u_0, s_i)} \right)}$
is the change in entropy associated with the change in state of the system.
If $u^r_s=u_{s_f-s}$ denotes the scale-reversed path associated with the stochastic process $u_s$, the entropy exchanged with the surrounding \mbox{medium~\cite{Lebowitz_1999, Chetrite_2008}}
\begin{eqnarray}
	\Delta S_{med}\left[u(\cdot)\right]  &=& -\ln \frac{\mathbb{P}\left[u_s^r = u(\cdot)\right]}{\mathbb{P}\left[u_s = u(\cdot) \right]},\\
	&=& \int_{s_i}^{s_f} \left[\dot{u}_s \frac{D^{(1)} - D'^{(2)}/2 }{D^{(2)}}\right]ds,
	\label{eq:Entropy_med}
\end{eqnarray}
which measures the irreversibility of the trajectory.
In Figure~\ref{fig:Action}(b) the comparison of $P(\Delta S_{med})$,
using eq.~(\ref{eq:Entropy_med}), indicates clear discrepancies for \mbox{K62:} besides a broader distribution, the relative frequency of entropy consuming trajectories 
is much lower.
For $\beta=0.044$ the model correctly reproduces the statistics of the negative and positive $\Delta S_{med}$.
For the total entropy variations, $\Delta S_{tot}$, the influence of $\beta$ on the integral fluctuation theorem (IFT) is presented in the Supplemental Material \cite{SM} \mbox{(Appendix F)}.
For $\beta=0.044$ the entropy variations are balanced by the IFT, whereas for $\beta=0$ this is not the case.
Supplemental Material \cite{SM} \mbox{(Appendix E)} show the correlation of of the two entropies.


\section{Cascade trajectories conditioned on $\Delta S_{med}$}
From Fig.~8 in the Supplemental Material \cite{SM} (Appendix H) we see that for experimental data the trajectories with $\Delta S_{med} <0$ contribute \mbox{up to $\approx 90\%$} to the heavy tailed statistics.
With increasing entropy-consumption a lower bound for $u_f$ becomes obvious. 
Examining the structure functions of these negative entropy trajectories negative scaling exponents are found as shown in Supplemental Material \cite{SM} \mbox{(Appendix I)}.

\begin{figure}
	\centering
	\includegraphics[width=0.28\textwidth]{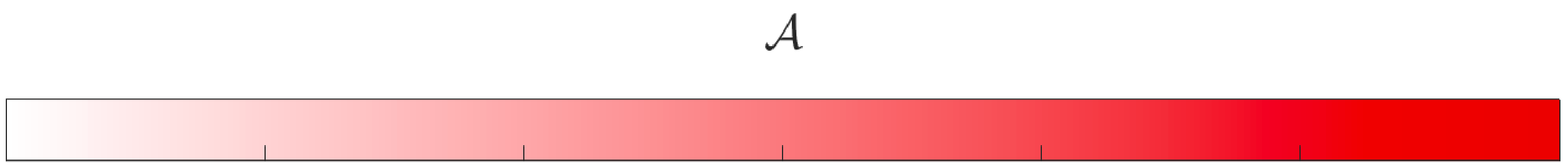}\\
	\includegraphics[width=0.291\textwidth]{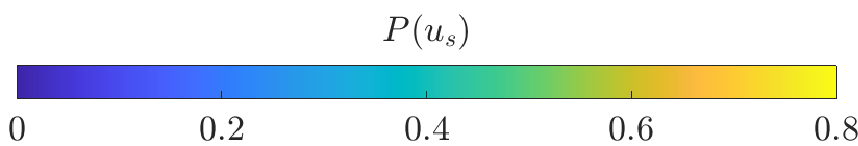}\\
	\includegraphics[width=0.215\textwidth]{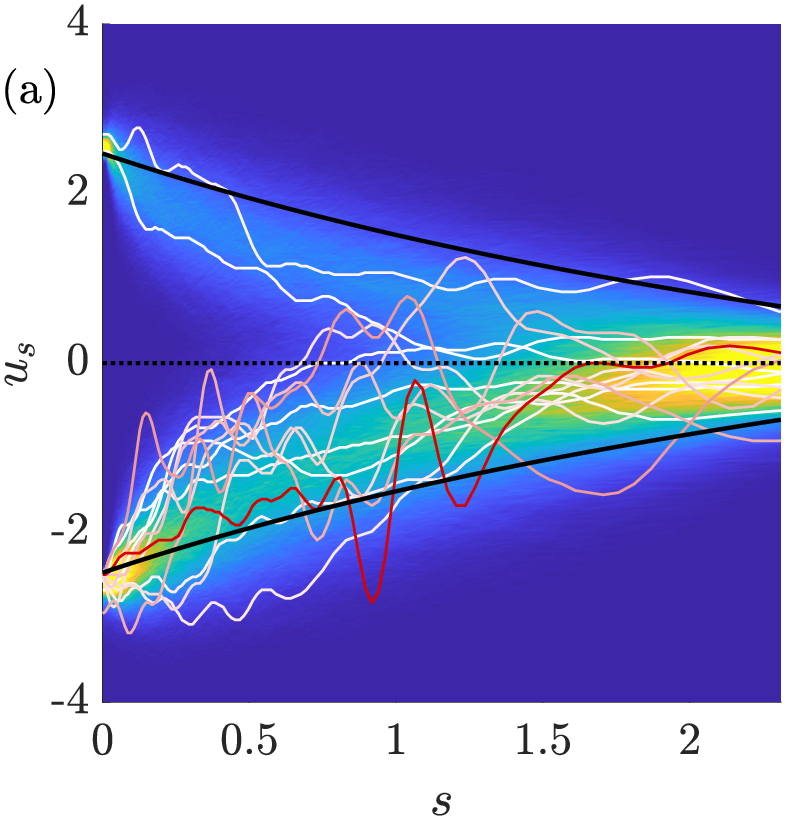}
	\includegraphics[width=0.215\textwidth]{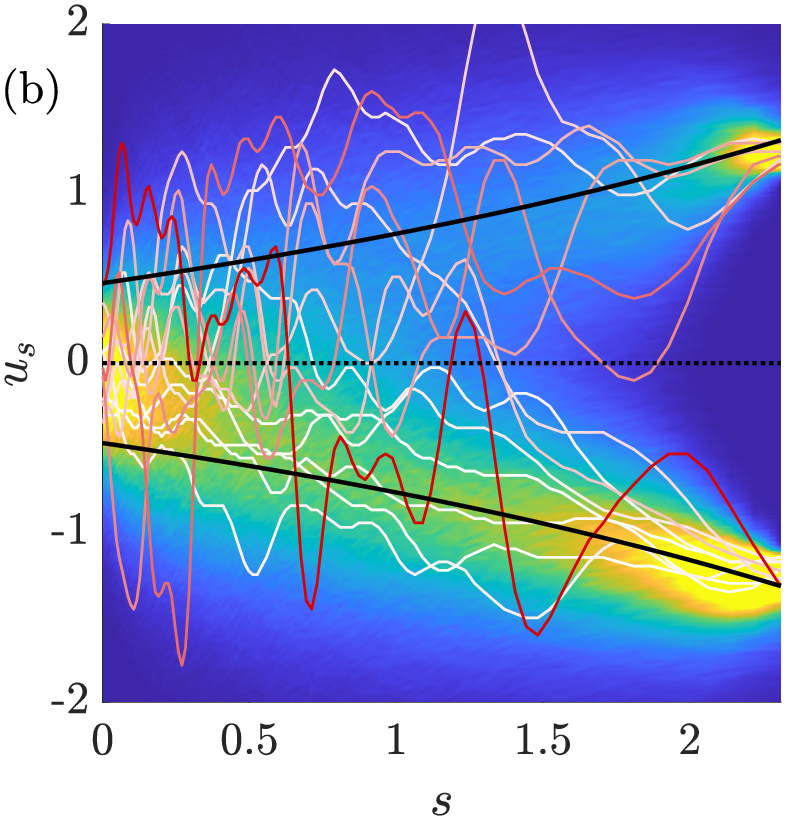}\\
	\includegraphics[width=0.215\textwidth]{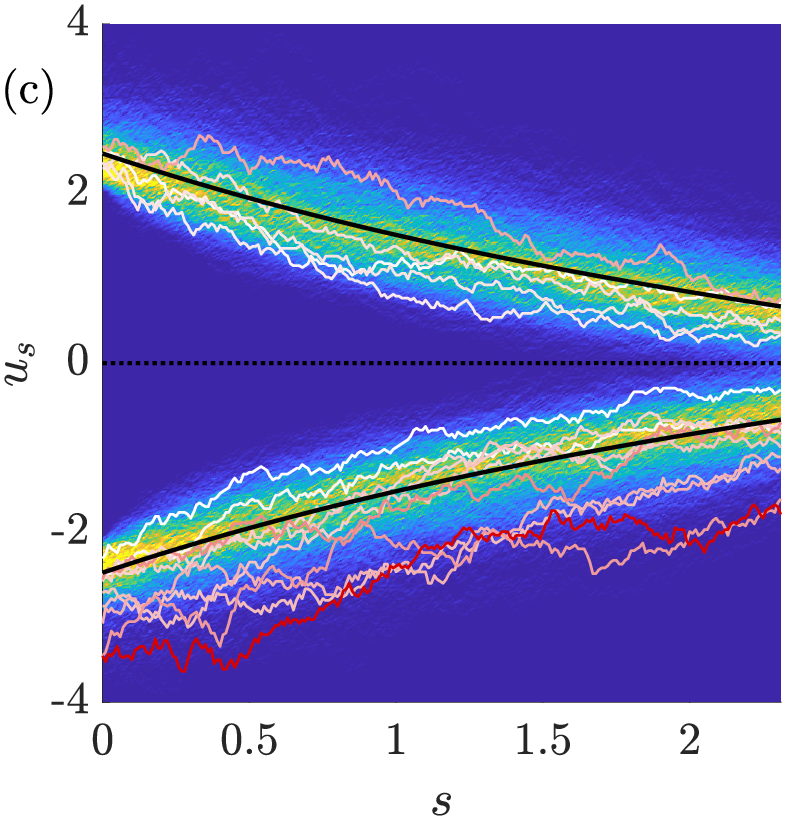}
	\includegraphics[width=0.215\textwidth]{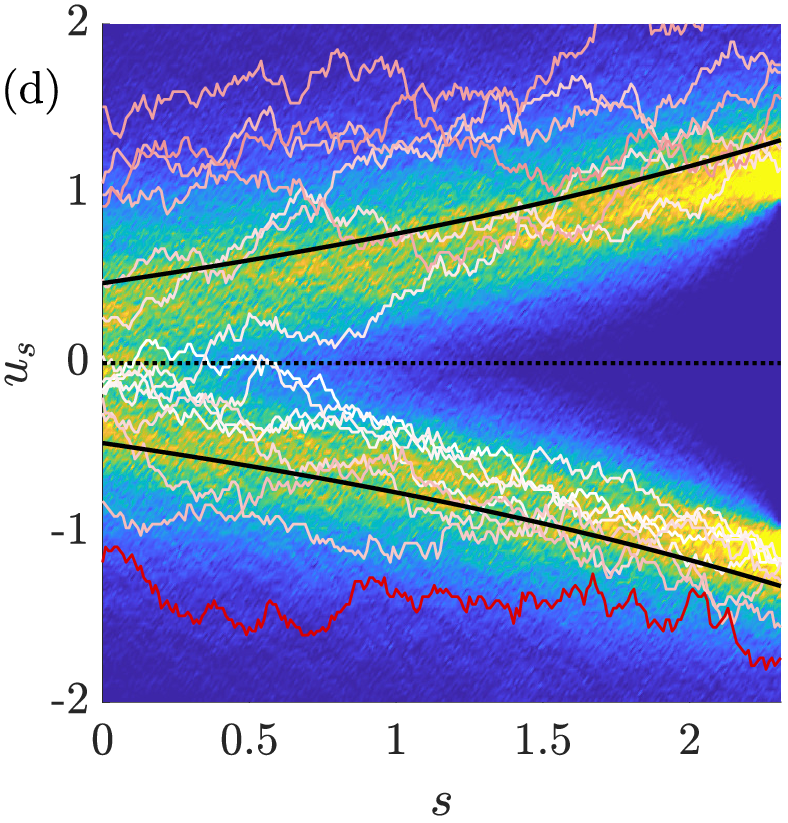}
	\caption{PDF of increment $u_s$ as a function of \mbox{scale $s$} (contours), conditioned on positive (a, c) and negative (b, d) entropy $\Delta S_{med}$, for the experimental (top row) and numerical data (bottom row) with \mbox{$\beta=0.044$}. The entropy value is from (a) to (d): $\Delta S_{med}=15, -5, 13$ and $-5$ (values selected so that the probability $P(\Delta S_{med})$ is approximately same (see Fig.~\ref{fig:Action}(b))). On top of the PDF are represented randomly chosen cascade trajectories characterized by the same entropy consumption (resp.\ production), color-coded using the value of their action $\mathcal{A}$ (increasing from white to red).
		The black solid resp.\ dotted lines represent the instanton trajectories for the diffusion model with $\beta=0.044$ and \mbox{$\beta=0$} computed by solving the variational problem given by eq.~(\ref{eq:actionvariationalproblementropy}).}
	\label{insta_exp_sim}
\end{figure}

In Fig.~\ref{insta_exp_sim}, contours of the PDF of increments conditioned on $\Delta S_{med}$ are presented 
for experimental and numerical data ($\beta=0.044$).
Conditioning trajectories reveal two distinct characteristics.
For entropy producing trajectories
\mbox{(see Fig.~\ref{insta_exp_sim}(a, c))} a splitting of the probabilities is seen for large scale increments $u_0$, which have the tendency to decay to typical increments with decreasing scale \mbox{$(u_0>u_f)|{\Delta S_{med}>0}$}, as it fits well to the traditional picture of turbulence \mbox{($u_r \approx r^{1/3} \approx e^{-s/3}$)}.
For entropy consuming trajectories \mbox{(see Fig.~\ref{insta_exp_sim}(b, d)),} the conditioned PDF markedly splits into two branches when approaching the Taylor scale $s_f$, corresponding typical increments at large scales develops into atypical increments at small scales \mbox{$(u_0<u_f)| {\Delta S_{med}<0}$}.
This splitting up can be interpreted as the dynamic fluctuations of small-scale intermittency.
%
In Fig.~\ref{insta_exp_sim} randomly selected individual conditioned trajectories, color-coded by the value of the corresponding action $\mathcal{A}$, are presented on top of the contours.
Trajectories with small action are located closer to the ridges of the PDF's,
while trajectories with large action depart further from it.
It can be noted that while the qualitative behavior described above is captured by both experimental data and diffusion model,
there are differences between the two subsist: experimental trajectories with large action have endpoints in the expected regions but tend to display large fluctuations away from the typical dynamics, while large action trajectories for the diffusion model fluctuate less but may start and end farther away from the high-probability regions.
Another difference 
is the asymmetry between negative and positive experimental trajectories, which is absent in the diffusion model.
Note that a symmetric distribution is assumed for the model at the initial condition 
 and this simple model
conserves skewness.

\section{Instanton formalism}
The conditioned PDF for increments has revealed the existence of two qualitatively different statistical dynamics.
Next we ask whether the ridge of the PDF
corresponds to an optimal trajectory given by the minimizers of the action functional, called \emph{instantons}.
If the contribution of such trajectories dominates the path integral, it is possible to compute the statistical properties of any observable using saddle-point approximations.
Even when fluctuations around these paths are important, instantons still correspond to most probable paths.

For known initial $u_0$ and final $u_f$ increments, instantons are trajectories which satisfy the variational problem
\begin{eqnarray}
	A(u_0, u_f) = \inf_{u} \left\{\mathcal{A}\left[u(\cdot)\right]\ |\ u(s_i)=u_0, u(s_f)=u_f \right\},
	\label{eq:actionvariationalproblem}
\end{eqnarray}
which can be solved using standard machinery of analytical mechanics.
Thus instantons are governed by \mbox{Hamilton equations}
\begin{eqnarray}
  H	  & =(\beta+\gamma u^2)p^2-(\alpha+2\gamma)up+\frac{\alpha+\gamma}{2}
\end{eqnarray}
with a generalized momentum $p$ and the relations
\begin{eqnarray}
	\frac{du}{ds} &=& +\frac{\partial H}{\partial p} =
	2 \left(\beta+\gamma u^2\right)p - \left(\alpha+2\gamma\right)u,\label{eq:hamiltoneq1}\\
	%
	\frac{dp}{ds} &=& -\frac{\partial H}{\partial u} =
	-2\gamma u p^2+\left(\alpha+2\gamma\right)p. \label{eq:hamiltoneq2}
\end{eqnarray}
%
Eq.~\eqref{eq:actionvariationalproblem} defines the boundary conditions $u(s_i)=u_0$, $u(s_f)=u_f$ for the Hamilton equations.

If the initial and/or final points are left unconstrained, the boundary conditions become
$p(s_i)=0$ and/or \mbox{$p(s_f)=0$}.
Here, it is natural to assume that the large-scale increment $u_0$ is not fixed, but random with the given probability distribution $P(u_0)$.
At the same time we shall leave $u_f$
unconstrained.
The probability of a path becomes weighted by $P(u_0)exp\left[-\mathcal{A}\left[u(\cdot)\right]\right]$
and the variational problem~eq.~(\ref{eq:actionvariationalproblem}) becomes $\inf_{u}\left\{\mathcal{A}\left[u(\cdot)\right] +f(u_0) \right\}$, where $f(u_0)=-\ln P(u_0)$.
A natural choice is a centered Gaussian distribution \mbox{$f(u_0)=u_0^2/(2\sigma^2)$}.
Minimizing this modified action yields the Hamilton equations~(\ref{eq:hamiltoneq1})--(\ref{eq:hamiltoneq2}) with boundary conditions $p(s_i)=f'(u_0)$ and \mbox{$p(s_f)=0$}.

The connection between
entropy 
and instanton formalism
is established by the statistics conditioned on $\Delta S_{med}\left[u(\cdot)\right]$.
The path integral formalism presented above still applies, restricting the integrals to the fields satisfying the imposed condition.
The conditioned instantons must satisfy 
\begin{equation}
	A(S) = \inf_{u} \left\{\mathcal{A}\left[u(\cdot)\right] +f(u_0) \ |\  \Delta S_{med}\left[u(\cdot)\right]=S \right\}.
	\label{eq:constrainedactionvp}
\end{equation}
In general, constrained variational problems are solved using Lagrange multipliers.
It is important to note that
while critical points of a constrained and relaxed variational problem coincide, the nature of the critical points may differ.
For the coefficients given by~eq.~(\ref{eq:betamodel})
the variational problem can be reduced to an unconstrained variational problem.
The entropy exchange defined in the Stratonovitch convention~\cite{GardinerBook} by eq.~(\ref{eq:Entropy_med}) can be computed analytically:
\begin{eqnarray}
	\Delta S_{med}\left[u(\cdot)\right] &=&- \frac{\alpha + 2\gamma}{2 \gamma} \ln \left( \frac{\beta+\gamma u_f^2}{\beta+\gamma u_0^2}\right). \label{eq:entropyexact}
\end{eqnarray}
Thus the entropy exchange is fixed by the value of the increments at the Taylor and integral scale ($u_f$ resp.\ $u_0$).
For a given entropy $S$,~eq.~(\ref{eq:entropyexact}) can be inverted $	\tilde{u}_f(S, u_0) = \sqrt{\frac{\beta}{\gamma}\left(e^{-\frac{2\gamma S}{\alpha+2\gamma}}-1\right)+u_0^2 e^{-\frac{2\gamma S}{\alpha+2\gamma}}}$,
and the constrained variational problem simplifies to
\begin{eqnarray}
	A(S) = \inf_{u_0} \left\{A(u_0, \pm\tilde{u}_f(S, u_0)) +f(u_0) \right\}. \label{eq:actionvariationalproblementropy}
\end{eqnarray}
To compute the conditioned instantons for $\beta \neq 0$
eq.~(\ref{eq:actionvariationalproblementropy}) can be solved numerically.
For the values of the entropy used in Fig.~\ref{insta_exp_sim}, two solutions with the same action are found in each case: one with positive $u_0$ and one with \mbox{negative $u_0$}.
For the data generated numerically from the diffusion model (Fig.~\ref{insta_exp_sim}(c, d)), the instantons match quite well the ridge of the PDF of increments and many individual trajectories are within the vicinity of the instanton.
In the K62 case ($\beta=0$), the variational problem can be solved analytically by exploiting the fact that the coupled ODE~\mbox{(\ref{eq:hamiltoneq1})--(\ref{eq:hamiltoneq2})} conserve $u\times p$, with $\frac{d}{ds} u \times p = u \frac{dp}{ds}  + p \frac{du}{ds} = 2 \beta p^2$.
We find that if the most probable increment at the integral scale is $u_0=0$, then the variational problem does not have a solution for $S \neq 0$ (see Supplemental Material \cite{SM}, Appendix G).
This explains why the K62 theory fails to capture the two distinct behaviors 
 shown 
in Fig.~\ref{insta_exp_sim}: in this model there is no preferred trajectory for a given entropy exchange.
In combination with the analysis of $P(\Delta S_{med})$ in Fig.~\ref{fig:Action}(b), this can be explained by the well known fact that K62 underestimates the frequency of large fluctuations on small scales~\cite{Frisch_1995, Anselmet_1984}.
Although significant differences between the instantons and the ridge of the PDF are evident for the experimental data \mbox{(Fig.~\ref{insta_exp_sim}(a, b))}, we emphasize that the overall behavior of entropy conditioned dynamics are reproduced qualitatively correct by the instantons for $\beta=0.044$.
Note, we have chosen a simplified stochastic process with constant coefficients $\alpha(s)$, $\beta(s)$
and $\gamma(s)$ to obtain this principal result.

\section{Conclusion}
The connection between the statistical quantity entropy and the structure related instanton formalism is worked out and may be referred to as an entropon.
The findings presented in this paper indicate that the entropy dependent dynamics through scales
are qualitatively reproduced by entropons, given by the analytical expression eq.~(\ref{eq:actionvariationalproblementropy}), if the K62 model is modified by adding the constant $\beta$ to the diffusion.
Note that the $\beta$ term breaks the scaling symmetry
and violates the conservation of $u \times p$. 
Further, the $\beta$ term has an important structural effect on entropons paths as the entropons for negative entropy confirm the surprising observation of zero probability of \mbox{$u_{f}\approx 0|{\Delta S_{med}< 0}$ for $\beta \neq 0$} (see Fig.~8 in the Supplemental Material \cite{SM} (Appendix H)
 and \mbox{Fig.~\ref{insta_exp_sim}(b, d)}).
Note these results could only be obtained due to the novel entropy constraints.
These constraints are expressed in terms of the fluctuation theorems, which indicate a statistical balance of negative and positive entropy events.
These findings provide a new perspective on the intermittency phenomenon, by pinpointing the trajectories to instantons responsible for the emergence of non-Gaussian statistics at small-scales, as has been proposed more generally for turbulence~\cite{Falkovich_1996,Grafke_2015}.
Entropons could be relevant for a wide range of problems in fluid dynamics as this connection provides an alternative way to common analysis incorporating the results into the statistical theory of turbulence and nonequilibrium thermodynamics.
Furthermore, the introduced instanton approach,  provides a test for models of the energy cascade that exceed the classical procedure to show how well anomalous scaling is reproduced.
In this work, all results including the underlying stochastic equations are deduced from the experimental data.
Perspective for future works include the computation of effective action and instantons, from the hydrodynamic equation, as has been considered for instance for problems in geostrophic turbulence~\cite{Bouchet_2018}.

\begin{acknowledgments}
We acknowledge financial support by Volkswagen Foundation (0324263), German Federal Ministry for Economic Affairs and Climate Action (03EE2031A), Laboratoire d’Excellence LANEF in Grenoble (ANR-10- LABX-51-01),
the European Union’s Horizon 2020 Research and Innovation Programme under the Marie Skłodowska-Curie Grant Agreement 753021. This publication was supported by a Subagreement from the Johns Hopkins University with funds provided by Grant No. 663054 from Simons Foundation (F. Bouchet). 
Its contents are solely the responsibility of the authors and do not necessarily represent the official views of Simons Foundation or the Johns Hopkins University.
We acknowledge helpful discussions with A. Girard, J. Friedrich, J. Ehrich, A. Engel, G. G\"ulker, S. Kharche, D. Nickelsen and \mbox{T. Wester}.
\end{acknowledgments}

\bibliographystyle{apsrev4-2}
\bibliography{biblio}


%
%
%
%

\end{document}